# The naturally designed spherical symmetry in the genetic code

Chi Ming Yang

*Physical Organic Chemistry and Chemical Biology, Nankai University, Tian Jin 300071, China*
*E-mail: yangchm@nankai.edu.cn*

ABSTRACT

In the present work, 16 genetic code doublets and their cognate amino acids in the genetic code are fitted into a polyhedron model. Based on the structural regularity in nucleobases, and by using a series of common-sense topological approaches to rearranging the Hamiltonian-type graph of the codon map, it is identified that the degeneracy of codons and the internal relation of the 20 amino acids within the genetic code are in agreement with the spherical and polyhedral symmetry of a quasi-28-gon, *i.e.*, icosikaioctagon. Hence, a quasi-central, quasi-polyhedral and rotational symmetry within the genetic code is described. Accordingly, the rotational symmetry of the numerical distribution of side-chain carbon atoms of the 20 amino acids and the side-chain skeleton atoms (carbon, nitrogen, oxygen and sulfur) of the 20 amino acids are presented in the framework of this quasi-28-gon model. Two evolutionary axes within the 20 standard amino acids are suggested.

## I. INTRODUCTION

The integrity, complexity and development mechanisms of the biological system form a puzzling subject. Wolpert and Lewis (1975) stated "The question of how these component parts (including the genetic networks) are organized into a complete control system for development is posed as a problem for future study". Particularly, the genetic code has been a conceptual challenge to all scientists since it was cracked down four decades ago. Towards understanding the origin and evolution of the genetic code, a few early landmarks include Crick's discussion about two fundamentally different theories of genetic code evolution, *i.e.*, "frozen accident theory" and the "stereochemical principle" or the "natural selection and co-evolution theory" (Crick, 1968; Wong, 1975), and an illuminative hypercycle principle from Eigen (Eigen, 1971; Eigen and Schuster, 1977; 1978; 1979; Eigen et al., 1981; Blomberg, 1997; Ycas, 1999).

Since Woese introduced a legend observation that the four RNA nucleobases, uracil (U), cytosine (C), adenine (A) and guanine (G) can be doubly ordered by ranking them according to their relative potentials as electron donors as well as their relative polarity/hydrophobicity (Woese et al., 1966; Woese, 1965; Woese, 1967), based on which the error minimization in the genetic code can be quantitatively measured (Haig and Hurst, 1999), it is now widely realized that genetic codons and amino acids can be shown to be related by chemical principles (Siemion and Stefanowicz, 1992; Rodin et al., 1993; Di Giulio et al., 1994; Cedergren and Miramontes, 1996; Di Giulio and Medugno, 1998; Knight and Landweber, 1998; Yarus, 1998; Knight et al., 1999; Yarus, 2000; Di Giulio, 2001; Di Giulio and Medugno, 2001). Presently, the internal regularity displayed by the genetic code has been recognized to include physicochemical property correlation, biosynthetic relation of amino acids and the non-random pattern of the genetic code.

Within the past few decades, accompanied with the codon origin and evolution of the genetic code which have fascinated scientists across numerous disciplinary fields, the inherent symmetry characteristics of the genetic code remains another unresolved but intriguing subject. Despite excellent efforts and innovative interdisciplinary approaches examining the unclear symmetry feature of the genetic code, however, a precise description of the symmetry characteristics inherent in the standard genetic code including, in particular, why the total number of standard amino acids is 20, still awaits a clear elucidation.

The 20 standard amino acids selected in the genetic code constitute a paradigm of complexity in Nature's integrity (Dufton, 1997; Davydov, 1998). Although biological importance of the code suggest a system complexity within the code, a multi-disciplinary in-depth examination of the code described here show its basic symmetric feature has great simplicity. In a previous paper (Yang, 2003) we presented a line of reasoning favoring the order of bases as UCGA succession in listing the genetic code. We used $sp^2$ N-atom number to rank nucleobases and the resulted genetic code shows a nice correlation between amino acid property and $sp^2$ N-numbers. The purpose of this work is to describe the newly identified symmetric three-dimensional relation of the 20 amino acids within the genetic code. This finding may have novel features, which are of considerable biological interest.

## II. METHOD AND RESULTS

1) An empirical stereo-electronic property of nucleobases for a rearranged genetic code

Chemical structures of the four nucleobases are mainly a six-membered ring, with or without infusion to a five-membered ring, in which all the heteroatoms (O, N) are in a conjugated position via their covalent bond linkages. Thus RNA nucleobases themselves display certain types of molecular structural regularity. I recently took one measure of this molecular regularity, covalent bonding hybrid of nitrogen atoms, as a determinative measure of chemical structural property to further discriminate among the mRNA nucleobase A, U, G and C. The $sp^2$ N-atom number in these nucleobases are 3 for A, 2 for G, 1 for C and 0 for U, respectively (Figure 1). If using this set of $sp^2$ N-atom numbers in the nucleobases as a determinative measure, a rearranged code is obtained in Table 1 and Figure 2, the latter is listed in a three-dimensional space.

Early studies demonstrated that a rearranged genetic code could sometimes be unexpectively informative (Grantham, 1980; Jiménez-Montaño et al., 1996; Jiménez-Montaño, 1999; D'Onofrio et al., 1999; Szathmary; 1999;Lehmann, 2000). RNA nucleobases listed in the succession of UCGA is coincidentally consistent with nucelobase hydrophilicity values from paper chromatography, which was invariably A<G<C<U (Weber and Lacey, 1978). Therefore, the genetic code listed in the order of AGCU, *i.e.*, increasing mononucleotide hydrophilicty in a correlation with increasing Woese's amino acid polar-requirement (Woese et al., 1966; Woese, 1965; Woese, 1967), was employed by Grantham (1980), Jimenez-Montano (Jiménez-Montaño et al., 1996; Jiménez-Montaño, 1999), D'Onofrio and co-workers (D'Onofrio et al., 1999) Szathmary (1999) and Lehmann (2000) and in their studies.

Number of $sp^2$ nitrogen atoms in RNA nucleic bases: A, G, C and U

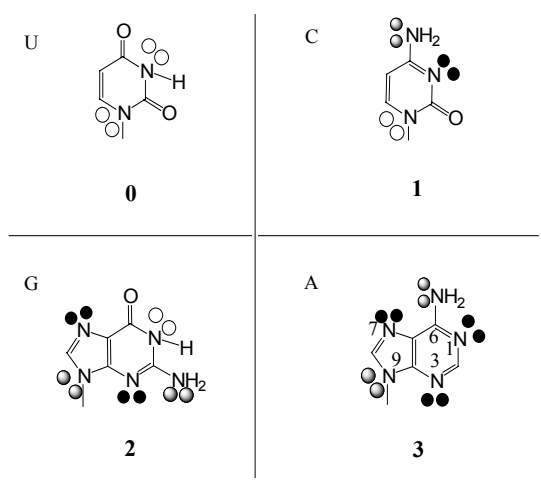

**Figure 1.** Nucleobases display unique chemical structure regularity. The four RNA nucleobases carry different number of $sp^2$ N atoms: 3 for A, 2 for G, 1 for C, to 0 for U, respectively. The nitrogen atom that carries one lone-electron-pair (LEP) ("• •") is a $sp^2$ hybrid nitrogen-atom.

**Table 1.**  A slightly rearranged codon map, where the four nucleobases forming tri-nucleotide codons are placed in the order of U, C, G and A.

| 1st letter) \ 2nd letter | U (aa) | | | C (aa) | | | G (aa) | | | A (aa) | | |
|---|---|---|---|---|---|---|---|---|---|---|---|---|
| U | UUU | (000) | F | UCU | (010) | S | UGU | (020) | C | UAU | (030) | Y |
|   | UUC | (001) | F | UCC | (011) | S | UGC | (021) | C | UAC | (031) | Y |
|   | UUG | (002) | L | UCG | (012) | S | UGG | (022) | W | UAG | (032) | Stop |
|   | UUA | (003) | L | UCA | (013) | S | UGA | (023) | Stop | UAA | (033) | Stop |
| C | CUU | (100) | L | **CCU** | **(110)** | **P** | CGU | (120) | R | CAU | (130) | H |
|   | CUC | (101) | L | **CCC** | **(111)** | **P** | CGC | (121) | R | CAC | (131) | H |
|   | CUG | (102) | L | **CCG** | **(112)** | **P** | CGG | (122) | R | CAG | (132) | Q |
|   | CUA | (103) | L | **CCA** | **(113)** | **P** | CGA | (123) | R | CAA | (133) | Q |
| G | **GUU** | **(200)** | **V** | **GCU** | **(210)** | **A** | **GGU** | **(220)** | **G** | GAU | (230) | D |
|   | **GUC** | **(201)** | **V** | **GCC** | **(211)** | **A** | **GGC** | **(221)** | **G** | GAC | (231) | D |
|   | **GUG** | **(202)** | **V** | **GCG** | **(212)** | **A** | **GGG** | **(222)** | **G** | GAG | (232) | E |
|   | **GUA** | **(203)** | **V** | **GCA** | **(213)** | **A** | **GGA** | **(223)** | **G** | GAA | (233) | E |
| A | AUU | (300) | I | **ACU** | **(310)** | **T** | AGU | (320) | S | AAU | (330) | N |
|   | AUC | (301) | I | **ACC** | **(311)** | **T** | AGC | (321) | S | AAC | (331) | N |
|   | AUG | (302) | M | **ACG** | **(312)** | **T** | AGG | (322) | R | *AAG* | *(332)* | ***K*** |
|   | AUA | (303) | I | **ACA** | **(313)** | **T** | AGA | (323) | R | *AAA* | *(333)* | ***K*** |

Amino acids (aa's) and their codons in blue and green color form a crossed intersection. Abbreviations of the 20 amino acids are represented by A(Ala), P(Pro), V(Val), G(Gly), T(Thr), S(Ser), L(Leu), R(Arg), D(Asp), E(Glu), M(Met), I(Ile), F(Phe), C(Cys), W(Trp), H(His), Q(Gln), N(Asn), K(Lys) and Y(Tyr).

2) A three-dimensional display of the 16 genetic code doublets in the rearranged genetic code

We now investigate the symmetry feature starting from a three-dimensional display of the rearranged genetic code. Given the 64 genetic codes comprising of 16 genetic code doublets, the strong group feature of the genetic code was initially recognized by Bertman and Jungck (1979), these genetic code doublets can be divided into two octets of completely degenerate and ambiguous coding dinucleotides. Findley and co-workers (1982) used a similar approach as that of Gatlin (1972), reasoned that the genetic code is a relation rather than a mapping by using empirical arguments processed within a group-theoretic framework (Findley et al., 1982; Gatlin, 1972). Findley discussed in particular the inherent symmetry as the even-order degeneracy constraint together with the odd-order degenerate codons in the genetic code.

Based on all of these work, the genetic code is now three-dimensionally presented in Figure 2, which more clearly display the internal relation between each two groups of the 16 genetic code doublets, with every line connecting two genetic code doublets, which vary by one base-letter from one to another.

To unravel the concealed symmetry feature within the genetic code, a series of topological approaches to rearranging the codon map is carried out. First, using a simple and common topological sense, the 3-D map in Figure 2a can be depicted by the following Hamiltonian graph in Figure 2b. In this graph, the internal relation between the 16 genetic code doublets are illustrated, each one of 16 genetic code doublets being connected with four other genetic code doublets, this connectedness corresponding to one base letter change between two neighboring genetic code doublets in the rearranged genetic code.

3) Rotational symmetry revealed from further topological transformation of the genetic code

To elucidate the hidden symmetry inherent in the genetic code, I further topologically rearrange the genetic code in a step depicted in Figure 3, to reach a closed spherical graph. Based on the discussion in the proceeding section and by placing the possible codon core on the bottom of the sphere, another way deciphering the 3-dimensional codon map is hence obtained in Figure 3b. A spherical shape in this display not only exhibits rotational symmetry and spherical feature of the genetic code system, but also graphically, in a visual sense, explains why the genetic code system has the capability of self-maintaining its integrity.

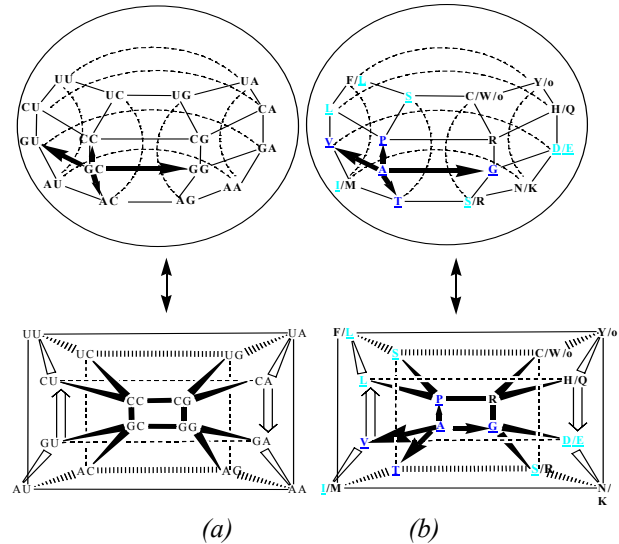

**Figure 2.** Reaching a Hamiltonian-type graphic display of the 16 genetic code doublets and their amino acids. Each vertex is connected with four other vertices and each vertex represents a genetic code doublet corresponding to its cognate amino acid(s): a) The 16 genetic code doublets of the rearranged genetic code in a three-dimensional space, whereby the four nucleobases are placed in the succession of UCGA. b) A three-dimensional display of the 20 amino acids corresponding to the 16 genetic code doublets. Four black arrows illustrate the core presumably formed at a certain evolution stage by five groups of genetic code doublets (amino acids are "A, P, V, G and T") (Yang, 2003). Amino acids in both blue and light-blue color are produced from Miller's prebiotic simulation experiment (Weber and Miller, 1981). "o" denotes "stop" codons. Abbreviations: 20 amino acids are represented by A(Ala), P(Pro), V(Val), G(Gly), T(Thr), S(Ser), L(Leu), R(Arg), D(Asp), E(Glu), M(Met), I(Ile), F(Phe), C(Cys), W(Trp), H(His), Q(Gln), N(Asn), K(Lys) and Y(Tyr); aa(amino acid).

(*Note 1:* We have previously re-classified the 20 amino acids into 5 groups of structures, according to their stereochemistry, which overlap precisely the 5 genetic code doublets in the crossed-intersection in a), it was suggested a codon core may have been formed at the primordial stage (16).)

[(*Note 2:* A Hamilton cycle is a cycle that includes each vertex exactly once (in other words, it is a spanning cycle). Since, if a map has a Hamilton cycle, then it can be four-colored (or 4-colorable) (Figure 3). A 4-colorable genetic code map is consistent with a requirement of four RNA bases in the whole genetic codes, in which the triplet (nucleotide) feature of codons reflects 3 dimension-character of each codon, as is displayed by each of the three base letters in every tri-nucleotide codon.) ]

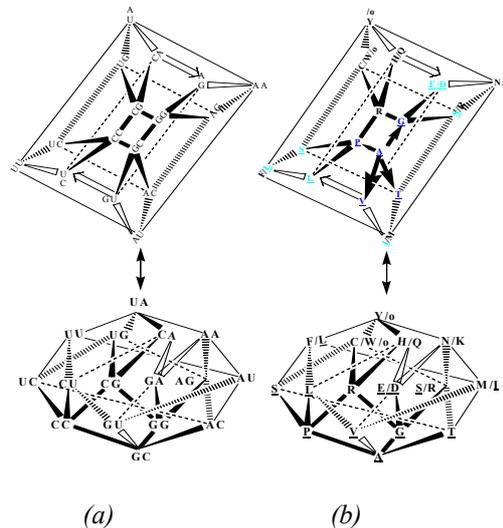

**Figure 3.** From Hamiltonian-type graph in Figure 2 to a closed spherical display which reveals the self-maintenance ability and the integrity of the genetic code. The 64 codons can be divided into 16 groups, half of that encode a single amino acid, having codon redundancy of 4. All the 3 sextets in degeneracy (each of the 3 amino acids encoded by 6 codons) are in the middle regions. Among the 5 amino acids forming a pyramid "A (P,V,G,T)", each amino acid is uniformly encoded by 4 codons from one genetic code doublet. The pyramid A (P,V,G,T) region is consistent with the early vision of primordial genetic code. Amino acids underlined are produced from Miller's simulation experiment (Weber and Miller, 1981). "o" means "stop" codon. The map shows that the more complex functionality an amino acid side-chain carries, the further away the amino acid is from Ala codons. If Ala is designated as the "Southern Pole", Tyr /o is the "Northern Pole". In this map, all the aromatic amino acids (Y, F, W, H) and stop codon "o" are in the same region, reminiscent of the current codon range expanding effort that is now being widely practiced in the "Tyr/o" codon region.

4) A comparison of the two three-dimensional map based on two genetic code table

Thus, when the arranged genetic code topologically presented in arrangement (II) on the basis of U, C, G and A succession, is compared with the commonly used genetic code presented in arrangement (I) whereby nucleobases are listed in a succession of U, C, A and G, it is evident that the new arrangement (II) from the rearranged genetic code (Table 1) can better display the high rotational symmetry in coding degeneracy, see Figure 4.

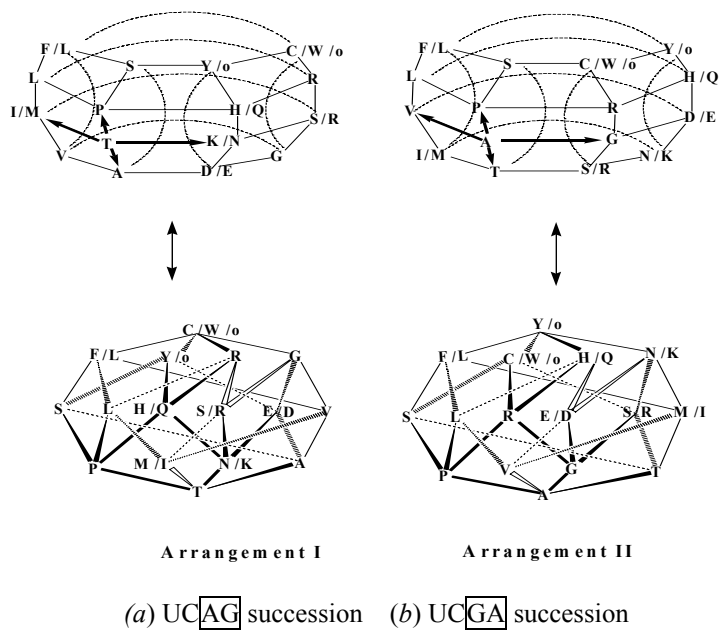

**Arrangement I**     **Arrangement II**

*(a)* UCAG succession     *(b)* UCGA succession

**Figure 4.** Two types of "world-map" arrangement of the 20 amino acids corresponding to two types of closed spherical representation of the 16 genetic code doublets: *I)* Nucleobases in the corresponding genetic code are placed in the succession UCAG; *II)* Nucleobases in the corresponding genetic code are placed in the succession UCGA

5) A polyhedral model (quasi-28-gon) helps reveal hidden symmetry of the genetic code

(1) Summarization of supersymmetry of the rearranged genetic code in the UCGA succession using a polyhedral model

Now, I select the code in arrangement II for a further discussion. Consider both the closed spherical feature with the newly identified rotational symmetrical feature of the code, the genetic code in arrangement II can be conveniently summarized by a polyhedron model in order to more definitely describe the internal symmetric relation of 16 genetic code doublets and their 20 cognate amino acids (Figure 5). In this model, symmetry in coding degeneracy can be described using a quasi-28-gon. In Table 2, one could amplify and generalize some of concepts.

The consequently elucidated amino-acid assignment and distribution around a quasi-28-gon comply with the general even-order degeneracy constraint (the fundamental degeneracy is of order 2 in genetic code), which is the basic symmetry as defined for the doubly degenerate codons. In addition to order-4 and order-6 degenerate codons in the genetic code, there are two sets of triply-degenerate codons, one of which maps onto Ile while the other maps onto Stop, and two nondegenerate codons, one of which maps onto Met while the other maps onto Trp. A quasi-28-gon helps clearly indicate that slight deviations from strict symmetry have occurred at the Y/o, C/W/o and M/I genetic code doublet positions. Despite these odd-order degenerate codons, nevertheless, the total number of amino acids at these positions remains C-symmetrical (Table 2). Notably, the "o" (stop) codons are not totally non-sense, but allow a counterbalance for numerical distribution of both amino-acid and side-chain C-atom along a presumed evolutionary axis (Table 2, 3 and Figure 7).

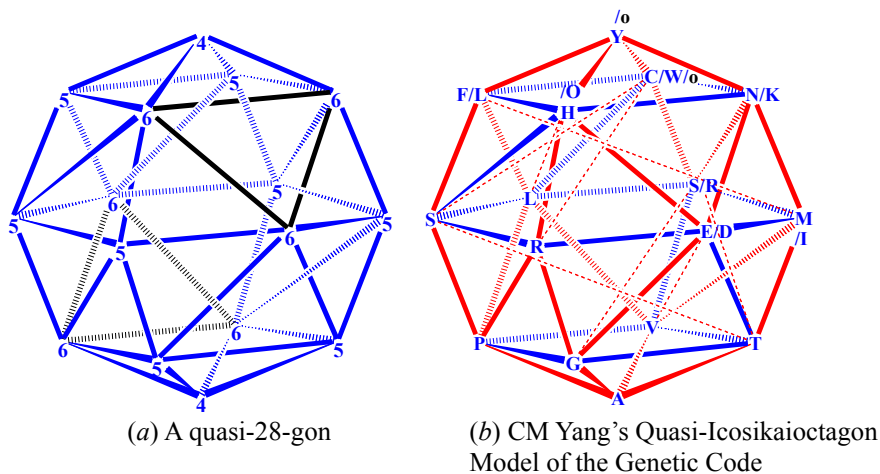

(*a*) A quasi-28-gon  (*b*) CM Yang's Quasi-Icosikaioctagon
Model of the Genetic Code

**Figure 5**.  Pattern of symmetry in the genetic code displayed in a polyhedron (quasi-icosikaioctagon or quasi-28-gon).  (*a*) The basic symmetry elements in a quasi-28-gon include 6 *C2* axes; 1 *C2* symmetric plan and 1 *I* (one symmetric center).  The number at each vertex indicates the number of edges.  (*b*) A solid geometrical illustration of the 3-dimensional codon map using a quasi-28-gon (Icosikaioctagon), displaying a highly symmetrical evolution of both the 20 amino acids and the 16 genetic code doublets.  The letter(s) at each vertex denotes amino acid(s) coded by one of the 16 genetic code doublets.  This map explains "why 20". The distribution of the 20 amino acids is symmetrical along a presumed evolutionary axis from Ala codons to Tyr codons: the number of the genetic code doublets (number of amino acids) at every stage starting from Ala codons to Tyr codons are: 1(1); 4(1,1,1,1); 6(1,1,1,2,2,2); 4(2,2,2,2) and 1(1,0), respectively.   In (*b*), block lines (both read and blue) are edges for polyhedron construction; red lines (both block and dotted) are for neighboring code-doublet connection.)

*Note 3:* For any map on a given surface, if the number of vertices of such a map is v, the number of arcs (or edges) is e, and the number of regions (or faces) is f, then, the number obtained from the expression v - e + f is denoted by *x*, and is termed the Euler characteristic of the surface.  For a sphere, *x* = 2.  Any polyhedron, which is both closed and convex, is termed a simple polyhedron.  Such a polyhedron may be continuously deformed into a sphere, and hence it follows that v - e + f = 2.  For Icosikaioctagon or 28-gon of the genetic code: *x* = v - e + f = 16 – e + 28 = 2; therefore, e = 42.

**Table 2.**  Symmetry in the degeneracy (numerical distribution) of the 20 amino acids and 16 genetic code doublets revealed by a 3-D code map (arrangement II in Figure 4) or the polyhedron model (Fig. 5).

| Trinucleotide codons and aa's | | Total number of aa's | Trinucleotide codons and aa's | | Total number of aa's | Trinucleotide codons and aa's | | Total number of aa's |
|---|---|---|---|---|---|---|---|---|
| | | | UAN | Tyr, "Stop" | 1 | | | |
| UUN | **Leu**, Phe | 4 | | | | CAN | His, Gln | 4 |
| UGN | Cys, Trp, "Stop" | | | | | AAN | Asn, Lys | |
| UCN | **Ser** | 3 | ⇑ | | | AGN | **Arg, Ser** | 6 |
| CUN | **Leu** | | | | | GAN | Asp, Glu | |
| CGN | **Arg** | | A presumed evolutionary axis of codons | | | AUN | Ile, Met, "Start" | |
| CCN | Pro | 2 | | | | GGN | Gly | 2 |
| GUN | Val | | | | | CAN | Thr | |
| | | | GCN | Ala | 1 | | | |

*Note 4:*  N = U, C, G and A.  Amino acids in black color are coded by codon from two different genetic code doublets, *i.e.*, Sextets (Quartets + Duets): Arg; Leu and Ser.   Degeneracy in the genetic code includes,
  *Even numbers:*
  3    Sextets (Quartets + Duets):   Arg; Leu; Ser;
  5    Quartets (Quadruplets):        Thr; Pro; Ala; Gly; Val;
  9    Duets:                         Lys; Asn; Gln; His; Glu; Asp; Tyr; Cys; Phe;
  *Odd numbers:*
  2    Triplets:                      Ile; terminators (STOP)
  2    Singlets:                      Met, Trp

(2) Cooperative symmetry of the numerical distribution of side-chain C-atoms of the 20 amino acids

Now we count the side-chain carbon atoms of amino acids at each genetic code doublets.  The cooperative symmetry in their distribution is then summarized in Table 3, and Figure 7.

**Table 3.** Symmetry in the numerical distribution of side-chain C-atoms of the 20 amino acids revealed a 3-D code map (arrangement II in Figure 4) or the polyhedron model (Fig. 5)

| Amino acids (C-atom number on side chains) | | Total number of C-atoms on side chains | Amino acids (C-atom number on side chains) | Total number of C-atoms on side chains | Amino acids (C-atom number on side chains) | | Total number of C-atoms on side chains |
|---|---|---|---|---|---|---|---|
| | | | UAN  Tyr(7), o | 7 | | | |
| **UUN** | **Leu(4)**, Phe(7) | 17 | ⇑ | | CAN | His(4),Gln(3) | 17 |
| AAN | Asn(2), Lys(4) | | | | UGN | Cys(1),Trp(9) | |
| **UCN** | **Ser(1)** | 13 | | | **AGN** | **Arg(4), Ser(1)** | 13 |
| GAN | Asp(2), Glu(3) | | A presumed evolutionary axis of codons | | **CUN** | **Leu(4)** | |
| AUN | Ile(4), Met(3), o | | | | **CGN** | **Arg(4)** | |
| CCN | Pro (3) | 5 | | | GGN | Gly(0) | 3 |
| ACN | Thr(2) | | | | GUN | Val(3) | |
| | | | GCN  Ala(1) | 1 | | | |

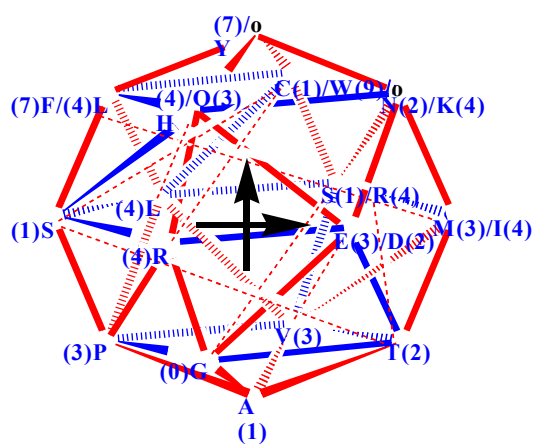

**Figure 6.** Number of carbon atoms on the side chains of amino acids. It is shown that any amino acids carrying a bigger number of side-chain C-atoms (>2) can be the summation of two amino acids carrying a smaller number of side-chain C-atoms, presumably consistent with a concerted stepwise coevolution of the canonical amino acids and their codons. Two possible evolutionary axes are indicated by arrows "↑" and "→".

Further information can be obtained from the quasi-28 model in Figure 7. We illustrate both the rotational symmetric feature of amino-acid coding degeneracy and symmetrical distribution of the sub-total number of C-atoms on side chains of amino-acid (s) at each vertex (genetic code doublet) based on a quasi-28-gon model. Moreover, results show that any amino acids carrying a bigger number of side-chain C-atoms (>2) can be the summation of two other amino acids carrying a smaller number of side-chain C-atoms, presumably consistent with a concerted stepwise evolution of the canonical amino acids and their codons---from simple to complex. Two evolutionary axes are thus proposed and indicated by arrows "↑" and "→".

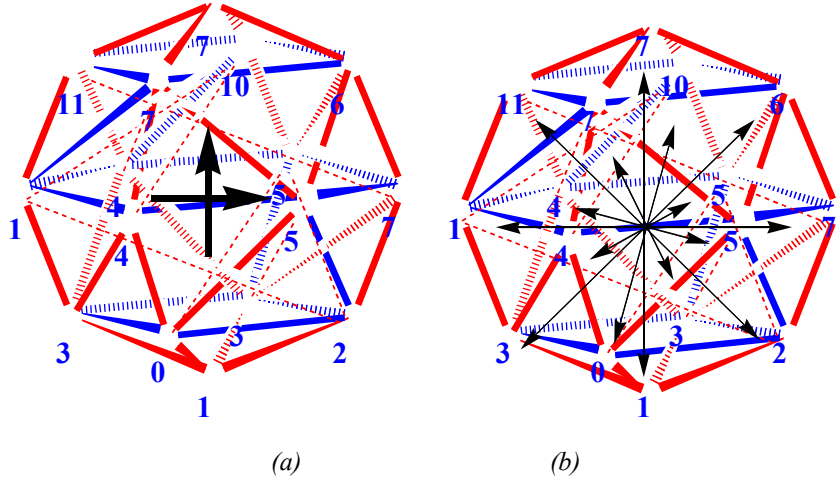

*(a)*          *(b)*

**Figure 7.** (*a*) Summarization of the sub-total number of C-atoms on side chains of amino acid(s) at each vertex (genetic code doublet) based on a quasi-28-gon model. Two proposed evolutionary axes are indicated by "↑" and "→". (*b*) *C2*-symmetry displayed in the sub-total number of side-chain C-atoms of amino acid(s) coded by two genetic code doublets at the opposite vertices, which are listed in groups of two perpendicular vertex-connection lines (+): 8, 8; 9, 9; 10, 10 and 9,13.

(3) Cooperative symmetry in the numerical distribution of side-chain skeleton (C, N, O, S) atoms of the 20 amino acids

The rotational symmetry, along the presumed evolutionary axis from Ala codons to Tyr codons, is also displayed as a cooperative symmetry in the numerical distribution of sub-total number of side-chain C(carbon)/N(nitrogen)/O(oxygen)/S(sulfur) atoms of the 20 amino acids based on a polyhedron model, see Table 4.

**Table 4.** Symmetry in the numerical distribution of sub-total number of side-chain C/N/O/S atoms of the 20 amino acids based on a polyhedron model.

| Amino acids (C/N/O/S-atom number on side chains) | Sub-total number of C/N/O/S-atoms on side chains | Amino acids (C/N/O/S-atom number on side chains) | Total number of C/N/O/S-atoms on side chains | Amino acids (C/N/O/S-atom number on side chains) | Sub-total number of C/N/O/S-atoms on side chains |
|---|---|---|---|---|---|
|  |  | UAN    Tyr(8), o | 8 |  |  |
| **UUN**    **Leu(4)**, Phe(7) | 11 |  |  | CAN    His(6),Gln(5) | 11 |
| AAN    Asn(4), Lys(5) | 9 |  |  | UGN    Cys(2),Trp(10), o | 12 |
| **UCN**    **Ser(2)**<br>**CUN**    **Leu(4)**<br>**CGN**    **Arg(7)** | 13 | ⇑<br>A presumed evolutionary axis of codons |  | **AGN**    **Arg(7), Ser(2)**<br>GAN    Asp(4), Glu(5)<br>AUN    Ile(4), Met(4), st | 26 |
| CCN    Pro (3)<br>ACN    Thr(3) | 6 |  |  | GGN    Gly(0)<br>GUN    Val(3) | 3 |
|  |  | GCN    Ala(1) | 1 |  |  |

## III. DISCUSSION

Early graphic approaches to understanding the genetic code include a genius display of the genetic code by employing a hypercube structural presentation by Jimenez-Montano et al. (1996), a codon graph group by Bertman and Jungck (1979), a geometric model study by Halitsky (1994) and a topological model by Karasev and Sorokin (1997). However, the precise symmetry features within the relation of the 20 amino acids have remained elusive. Now, if one uses a polyhedron to illustrate the genetic code table 1, we can have a clearer view of the spherical relation within the 16 genetic code doublets and the 20 amino acids in the genetic code. Upon an integrated approach to looking at the geometry, symmetry, chemistry and topology of the genetic code, it was identified that both the distribution of the 20 amino acids and degeneracy of codons are in agreement with the cooperative symmetry of a polyhedron 28-gon (or icosikaioctagon, Figure 5).

From this polyhedron model, we can find that the symmetric distribution of the 20 amino acids around the 28-gon excellently fits Rumer's regularity and explain the arithmetical regularity discovered by Shcherbak (1993). Both the amino acid in Group IV (degeneracy 4) and the rest of the amino acid within Quasi-group III-II-I (degeneracy 1; 2 and 3) are occupying half of the 28-gon (Figure 6).

Therefore, a general symmetric feature of the code described by our spherical and polyhedral approach together with two evolutionary axes is not only consistent with previous work, but also in an excellent agreement with the recent Trivonov's proposal that Ala could be the first codonic amino acid in the code that has been reached previously by quite other reasoning (Trifonov and Bettecken, 1997; Trifonov, 2000). The conclusion that a few amino acids and their codons are initially formed in the primordial stage was also envisaged by numerous previous works (Siemion and Stefanowicz, 1992; Rodin et al., 1993; Di Giulio et al., 1994; Cedergren and Miramontes, 1996; Di Giulio and Medugno, 1998; Knight and Landweber, 1998; Yarus, 1998; Knight et al., 1999; Yarus, 2000; Di Giulio, 2001; Di Giulio and Medugno, 2001).

The degeneracy in the standard genetic code has been shown to be ambiguous under certain condition, that is, more than one amino acid appears to correspond to the same codon, nevertheless, it has long been postulated that distribution of the total number of amino acids and "why 20" shall follow a symmetric pattern with the codon table (Weber and Miller, 1981). So far, symmetry and coding properties of genetic code has received tremendous attention (Portelli and Portelli, 1985; Shcherbak, 1988; Chipens et al., 1989; Shcherbak, 1989; Chipens, 1991a; 1991b; Hornos and Hornos, 1993; Arquest and Michel, 1996; Koch and Lehmann, 1997; Bashford et al., 1998; Frappat et al., 1999; Hornos et al., 1999; Bashford and Jarvis, 2000; Frappat et al., 2001; Lacan and Michel, 2001; Balakrishnan, 2002), and the pattern and symmetries inherent in the genetic code has been subjected to a wide variety of investigation approaches, including group theoretical analysis such as Lie algebra (Halitsky, 1994; Bashford et al., 1998; Hornos et al., 1999; Bashford and Jarvis, 2000; Balakrishnan, 2002;), hypercube structural presentation (Jimenez-Montano et al., 1996)), codon graph group (Bertman and Jungck, 1979) and geometric model study (Halitsky, 1994; Karasev and Sorokin, 1997). A codon ring as a theoretical model showed that the genetic code is almost an example of a Gray code (Swanson, 1984). The genetic code was more recently shown as a binary code, determined by Golden mean through the unity of the binary-code tree and the Farey tree (Rakocevic and Jokic, 1996).

Based on the quasi-28-gon model, it is identified herein not only the codon degeneracy is rotationally symmetric along a presumed evolutionary axis from Ala to Tyr codons, but also the side-chain C-atom numbers as well as the side-chain C/N/O/S atom-numbers are distributed in a rotationally symmetric way along this evolutionary axis (Table 4). This finding may be of interest to the current code expanding effort (Döring et al., 2001; Wang et al., 2001 ).

Although it has been known the genetic code is optimized and fixed (Freeland and Hurst, 1998; Freeland et al., 2000), its contents, in particular, the fundamental but unclear physicochemical property of 20 amino acids together with the distribution of amino acids in the genetic code has since inspired tremendous amount of excellent research and speculation including from Weber and Miller (1981), Swanson (1984), Luo (1989), Shcherbak (1993), Dufton (1997) and Davydov (1998). When Rakočević and Jokic (1996) suggested that the genetic code is determined by Golden mean, is was identified that atom number balance in amino acids are directed by Golden mean route, also directed by the double-triple system of amino acids, as well as by two classes of enzymes aminoacyl-t-RNA synthetases. Rosen (1999) described that the codon range numbers that follow for the 20 amino acids are shown to be given by linear Diophantine and explicit molecular content equations in the number of carbon, nitrogen, oxygen and sulfur atoms in each amino acid, suggesting that the universal genetic code that associates codons and amino acids is expressed in a precise way by purely physical molecular content relations.

To the best of my knowledge, it has for the first time been identified that the polyhedral symmetries as a quasi-28-gon are inherent in the genetic code. The spherical and polyhedral feature in the genetic code presented here is aimed to understand not just what now is, but the ways what now is might plausibly be expected to have arisen (Kauffman, 1993).

In summary, the rotational and spherical symmetry of codon-amino acid relation are displayed as

follows:
a. According to our quasi-28-gon model, there is rotational symmetry and a deviated symmetric center within the code, but this symmetry is slightly destroyed by evolution in the code;
b. The long-conjured co-operative symmetry character in the genetic code can be more precisely described by using a quasi-28-gon;
c. The circular property of the code and hypercycle theory of the genetic code can be further explored by using the quasi-28-gon model;
d. Amino acids distribution within the genetic code is symmetric along a possible evolutionary axis; This symmetry may have been established in the early evolutionary stage, in which A, P, V, G and T may have formed a possible core (Yang, 2003), consistent with a notion that Ala codons may be the first amino acid and codons;)
e. The 16 genetic code doublets from the 64 codons can be divided into 5 stages along a presumed evolutionary axis (Figure 6). Those 5 stages contain 1 group, 4 groups, 6 groups, 4 groups, 1 group of genetic code doublets, respectively. The numbers of genetic code doublets and the number of amino acids encoded at every stage are: 1(1); 4(1,1,1,1); 6(1,1,1,2,2,2); 4(2,2,2); 1(1,0). Exactly half of the 16 genetic code doublets are encoding 2 amino acids within each genetic code doublets except at Tyr position, and occupying half of the 3-D codon map. Alternatively, as revealed in the polyhedral model, the 20 standard amino acids and their codons would be stepwise evolutionarily divided into four groups, which may reflect their chronological order: A > P, V, G, T > S, L, R, D, E, M, I > F, C, W, H, Q, N, K > Y. Therefore, the polyhedron model, quasi-28-gon, supports our early suggestion that an early "frozen core" may have been formed in the primordial stage to attain 5 codon quartets for A and P, V, G and T, which may have inherent tendency in shaping the rest of genetic code (Yang, 2003).
f. Results from the quasi-28-gon model are in agreement with a coevolution theory of the genetic code. Since code and amino acids are biochemically and/or stereochemically linked, as evident from Figure 4b and 5b, amino acid evolutionary trend along the presumed evolutionary axis in this model is consistent with chemical evolution, e.g., from D/E to Q/N; F to Y and S to C;
g. Given the current interest in code range expansion (Döring et al., 2001; Wang et al., 2001 ), the quasi-28-gon model for the code reported herein is especially worthy of our attention. The polyhedral symmetry corresponding to quasi-28-gon helps understand the logic underlying why incorporating Tyr-derivatives (O-methyl-L-tyrosine, photocrosslinking amino acid) into amber codon (UAG) will not break the rotational symmetry along the presumed evolutionary axis, from Ala codons to Tyr codons (Figure *3b*, *5b*), within the genetic code. Similarly, when amber codons in the methyltransferase genes of certain archaea encode pyrrolysine as the 22$^{nd}$ amino acid, the rotational symmetry is little affected. However, when selenocysteine, identified as the 21st amino acid (in 1986), was directly encoded by UGA, which otherwise usually specifies translation termination (stop codon), actually breaks the rotational symmetry along the presumed evolutionary axis within the genetic code.

Since spherical symmetry is always associated with polyhedral symmetry, therefore, the polyhedron relation of the genetic code with features of rotational symmetry is not merely a natural accident. Icosahedral symmetry, which is a type of rotational symmetry, is regularly encountered, described as *Ih* symmetry. This is a common structure in virus coats, or capsids. A number of viruses such as poliovirus with its genome of RNA, have a protein capsid with a structural symmetry of icosahedron (Racaniello, 1996; Vargas et al., 1999; Lanzavecchia et al., 2002). The quasi-28-gon model may immediately suggest, in addition to the evolutionary logic currently tested by code-expanding effort, a possible mechanism in RNA translation. Therefore, it may not escape our notice that the quasi-28-gon symmetry in the genetic code with two "poles" as Ala codons and Tyr codons may have multiple biological implications, including the complex protein-synthesis mechanism which imposes another challenging problem for investigation (Liljenstrom and Blomberg, 1987; Krakauer and Jansen, 2002).

ACKNOWLEDGMENTS: The author is grateful to Dr. K. Chen, Mrs. Y. T. Li, Y. Bai, S. Z. Luo, T. Huang and Ms. X. F. Zhao for technical assistance.

ABBREVIATIONS: U, uracil; C, cytosine; A, adenine; G, guanine. aa, amino acid; A(Ala), P(Pro), V(Val), G(Gly), T(Thr), S(Ser), L(Leu), R(Arg), D(Asp), E(Glu), M(Met), I(Ile), F(Phe), C(Cys), W(Trp), H(His), Q(Gln), N(Asn), K(Lys) and Y(Tyr).

FOOTNOTES:
‡ Corresponding author: Fax: + 86 22 2350 3863.   E-mail address: yangchm@nankai.edu.cn

*(June 2003)*